\documentclass[aps,prl,twocolumn,amsmath,amssymb,reprint,superscriptaddress,preprintnumbers,nofootinbib]{revtex4}

\usepackage{enumitem}
\usepackage{graphicx}
\usepackage{slashed}

\newcommand{\sign}{\mathop{\mathrm{sign}}\nolimits}
\newcommand{\NP}{\slashed{\mathcal{P}}}
\newcommand{\BS}{\mathbb{S}}
\begin{document}

%\preprint{}

%\preprint{DESY~20--033\hspace{13.5cm}ISSN 0418--9833\phantom{XXXXXX}}
%\preprint{October 2020\hspace{17.3cm}}

\title{Four-loop anomalous dimension of flavor non-singlet twist-two operator of general Lorentz spin in QCD: $\mathbf{\zeta(3)}$ term}


\author{B.~A.~Kniehl}
\email[]{kniehl@desy.de}
%\thanks{}
\affiliation{Department of Physics, University of California at San Diego, 9500 Gilman Drive, La Jolla, CA 92093-0112, USA}
\affiliation{{II.} Institut f{\"u}r Theoretische Physik,
Universit{\"a}t Hamburg,
Luruper Chaussee 149, 22761 Hamburg, Germany}
\altaffiliation{}
\author{V.~N.~Velizhanin}
\email[]{vitaly.velizhanin@desy.de}
%\thanks{}
\affiliation{{II.} Institut f{\"u}r Theoretische Physik,
Universit{\"a}t Hamburg,
Luruper Chaussee 149, 22761 Hamburg, Germany}
%\affiliation{Theoretical Physics Division, NRC ``Kurchatov Institute,''
%Petersburg Nuclear Physics Institute,
%Orlova Roscha, Gatchina, 188300 St.~Petersburg, Russia}

\date{\today}

\begin{abstract}
  We consider the anomalous dimension of the flavor non-singlet twist-two quark operator of arbitrary Lorentz spin $N$ at four loops in QCD and construct its contribution proportional to $\zeta(3)$ in analytic form by applying advanced methods of number theory on the available knowledge of low-$N$ moments.
  In conjunction with similar results on the $\zeta(5)$ and $\zeta(4)$ contributions, this completes our knowledge of the transcendental part of the considered anomalous dimension.
This also provides important constraints on the as-yet elusive all-$N$ form of the rational part.
Via Mellin transformation, we thus obtain the exact functional form in $x$ of the respective piece of the non-singlet Dokshitzer--Gribov--Lipatov--Altarelli--Parisi splitting function at four loops.
This allows us to appreciably reduce the theoretical uncertainty in the approximation of that splitting function otherwise amenable from the first few low-$N$ moments.
\end{abstract}

%\keywords{}

\maketitle

%\section{}
%\subsection{}
%\subsubsection{}

In the framework of the collinear parton model of QCD \cite{Bjorken:1969ja}, with partons being gluons and quasi-massless quarks, parton distribution functions (PDFs) are crucial ingredients for the theoretical prediction of any production process in high-energy collisions involving hadrons in the initial state, including hadroproduction, photoproduction, and deep-inelastic scattering.
PDFs cannot yet be derived from first principles of QCD.
Exploiting their universality guaranteed by the factorization theorem \cite{Collins:1989gx}, however, they can be reliably determined through global fits to measured production cross sections.
By the same token, their dependencies on the typical energy scale of the scattering process are pinned down by the Dokshitzer--Gribov--Lipatov--Altarelli--Parisi (DGLAP) evolution equations \cite{Gribov:1972ri,Gribov:1972rt,Altarelli:1977zs,Dokshitzer:1977sg}.
The DGLAP evolution kernels are expressed in terms of the so-called splitting functions $P_{ij}(x)$, which, in a way, measure the probabilities of parton $j$ to semi-inclusively branch to parton $i$ carrying the fraction $x$ of the longitudinal momentum of $j$ plus one or more other partons, whose phase space is integrated over.
While PDFs are genuinely nonperturbative objects, splitting functions can be calculated order by order in perturbation theory,
\begin{equation}
P_{ij}(x)=\sum_{n=0}^{\infty}a_s^{n+1}P_{ij}^{(n)}(x)\,,
\end{equation}
where $a_s=\alpha_s(\mu)/(4\pi)$ with $\alpha_s(\mu)$ being the strong-coupling constant.
This may be done by directly evaluating the Feynman diagrams of the parton splittings including their radiative corrections \cite{Altarelli:1977zs}.
An alternative, actually more powerful, approach relies on the fact that the splitting functions are related via Mellin transformation,
\begin{equation}
\gamma_{ij}(N)=-\int_0^1 dx\ x^{N-1} P_{ij}(x)\,,
\end{equation}
to the anomalous dimensions
\begin{equation}
\gamma_{ij}(N)=\sum_{n=0}^{\infty}a_s^{n+1}\gamma_{ij}^{(n)}(N)\,,
\end{equation}
of appropriately defined local composite operators of twist two and Lorentz spin $N$, constructed from quark fields, gluon field strength tensors, and a definite number of covariant derivatives, namely $N-1$ for quark operators and $N-2$ for gluon operators \cite{Dokshitzer:1977sg}.

The leading-order terms, $P_{ij}^{(0)}(x)$ and $\gamma_{ij}^{(0)}(N)$, arise from tree-level and one-loop diagrams \cite{Gribov:1972ri,Gribov:1972rt,Altarelli:1977zs,Dokshitzer:1977sg,Gross:1973ju,Gross:1974cs}, respectively, and so on.
The next-to-leading-order \cite{Floratos:1977au,GonzalezArroyo:1979df,Floratos:1978ny,GonzalezArroyo:1979he,Gonzalez-Arroyo:1979kjx,Curci:1980uw,Furmanski:1980cm} and next-to-next-to-leading-order results \cite{Larin:1996wd,Moch:2004pa,Vogt:2004mw} are fully known analytically.
At next-to-next-to-next-to-leading order, only partial results are available so far \cite{Gracey:1994nn,Velizhanin:2011es,Velizhanin:2014fua,Davies:2016jie,Moch:2017uml,Davies:2017hyl,Moch:2018wjh,Das:2020adl,Moch:2021qrk,Falcioni:2023luc,Gehrmann:2023cqm,Falcioni:2023vqq,Falcioni:2023tzp,Moch:2023tdj,Gehrmann:2023iah,Falcioni:2024xyt,Falcioni:2024xav,Falcioni:2024qpd}.
The complexity of these four-loop computations rapidly grows with $N$, so that the quest for high-$N$ moments has come to a grinding halt.
Further progress is limited by the want of more powerful computers.

The lack of knowledge of $\gamma_{ij}^{(3)}(N)$ for general value of $N$ prevents us from obtaining $P_{ij}^{(3)}(x)$ in exact analytic form, so that we have to resort to approximations \cite{Larin:1996wd,Moch:2017uml,Moch:2023tdj}, whose errors are difficult to control and bound to be particularly large at low values of $x$ (see, {\it e.g.}, Figs.~5 and 6 in Ref.~\cite{Moch:2017uml}).
On the other hand, present and future high-precision measurements at the CERN Large Hadron Collider (LHC) and the BNL Electron-Ion Collider (EIC) dramatically boost the benchmarks for the uncertainties in the PDFs required to meet this challenge on the theoretical side.
This strongly motivates us to recover all-$N$ results for $\gamma_{ij}^{(3)}(N)$ as much as possible.

Given the PDFs of $f$-flavored quarks and antiquarks and the gluon, $q_f$, $\bar{q}_f$, and $g$, is advantageous to organize the quark sector in terms of flavor asymmetries $q_{\mathrm{ns},ff^\prime}^{\pm}=(q_f\pm \bar{q}_f)-(q_{f^\prime}\pm \bar{q}_{f^\prime})$, flavor non-singlet valence $q_{\mathrm{ns}}^{\mathrm{v}}=\sum_f(q_f-\bar{q}_f)$ and flavor pure singlet $q_{\mathrm{s}}=\sum_f(q_f+\bar{q}_f)$.
Then the DGLAP evolution proceeds separately for $q_{\mathrm{ns},ff^\prime}^{\pm}$ and $q_{\mathrm{ns}}^{\mathrm{v}}$, leaving a $2\times2$ matrix equation for $q_{\mathrm{s}}$ and $g$.
This involves seven distinct splitting functions, $P_{\mathrm{ns}}^{\pm}$, $P_{\mathrm{ns}}^{\mathrm{v}}$, $P_{qq}$, $P_{qg}$, $P_{gq}$, $P_{gg}$, which are mirrored by respective anomalous dimensions in Mellin space.
In this letter, we focus our attention on $\gamma_{\mathrm{ns}}^{\pm(3)}(N)$.
Operator product expansion reveals that $\gamma_{\mathrm{ns}}^{\pm(3)}(N)$ vanishes for odd and even values of $N$, respectively, so that it is convenient to merge them into the single quantity $\gamma_{\mathrm{ns}}^{(3)}(N)$ using the sign factor $\varepsilon=(-1)^N$.
This is known to break down in rational and transcendental parts as
\begin{equation}
\gamma_{\mathrm{ns}}^{(3)}(N)=
\gamma_{\mathrm{rat}}^{(3)}(N)
+\zeta_3\gamma_{\zeta_3}^{(3)}(N)
+\zeta_4\gamma_{\zeta_4}^{(3)}(N)
+\zeta_5\gamma_{\zeta_5}^{(3)}(N)\,,
\label{eq:zeta3}
\end{equation}
where $\zeta_k=\zeta(k)$ is Riemann's zeta function.
A term proportional to $\zeta_2=\pi^2/6$ is prohibited by the no-$\pi^2$ theorem \cite{Davies:2017hyl,Jamin:2017mul,Baikov:2018wgs}.
Presently, $\gamma_{\zeta_4}^{(3)}(N)$ \cite{Davies:2017hyl} and $\gamma_{\zeta_5}^{(3)}(N)$ \cite{Moch:2017uml} are known for all $N$, while $\gamma_{\mathrm{rat}}^{(3)}(N)$ and $\gamma_{\zeta_3}^{(3)}(N)$ are only available for $N=1,\ldots,16$ \cite{Velizhanin:2011es,Velizhanin:2014fua,Moch:2017uml}.
Furthermore, the all-$N$ results are known for the $n_f^3$ \cite{Gracey:1994nn}, $n_f^2$ \cite{Davies:2016jie}, and QED-like $C_F^3n_f$ contributions to $\gamma_{\mathrm{ns}}^{(3)}(N)$, its large-$n_c$ limit in SU($n_c$) color gauge theory \cite{Moch:2017uml}, and its counterpart in $\mathcal{N}=4$ supersymmetric Yang--Mills theory (SYM) \cite{Beisert:2006ez,Bajnok:2008qj}.
In the following, we will analytically reconstruct the all-$N$ result for $\gamma_{\zeta_3}^{(3)}(N)$ utilizing advanced techniques of number theory in combination with an understanding of the appearing classes of special functions, conjectured theorems, and educated guesses.
We will not inject the $C_F^3n_f$ \cite{Gehrmann:2023iah}, large-$n_c$ \cite{Moch:2017uml}, and $\mathcal{N}=4$ SYM \cite{Beisert:2006ez,Bajnok:2008qj} results, but rather use them for cross checks.

Detailed inspection of the known all-$N$ results for $\gamma_{ij}^{(n)}(N)$, at $n+1$ loops, and parts thereof reveals that the special functions are exhausted by the nested harmonic sums \cite{Vermaseren:1998uu},
\begin{equation} 
S_{m_1\ldots m_M}(N)=\sum^{N}_{i=1} \frac{[\sign(m_1)]^{i}}{i^{\vert m_1\vert}}
\,S_{m_2\ldots m_M}(i)\,,
\end{equation}
where $S(N)=1$ and $m_j\in\mathbb{Z}$, with weight $w=\sum_{j=1}^M|m_j|\le2n+1$, and that they appear as linear combinations with rational coefficients.
At weight $w$, there are $2\times3^{w-1}$ such harmonic sums.
This set of functions may serve as a basis for an ansatz for the analytic reconstruction of some as-yet unknown $\gamma_{ij}^{(n)}(N)$ for all $N$.
Unfortunately, the dimension of this function space rapidly grows with $n$, being $2,18,162,1458,\ldots$ for $n=0,1,2,3,\ldots$.
Notice that $\zeta_k$ has $w=k$, each power of $n_f$ counts as $w=1$, and their appearances as overall factors correspondingly reduce the maximum weight of the basis.
This explains why $\gamma_{\zeta_5}^{(3)}(N)$ \cite{Moch:2017uml}, $\gamma_{\zeta_4}^{(3)}(N)$ \cite{Davies:2017hyl}, and the terms proportional to $n_f^3$ \cite{Gracey:1994nn} and $n_f^2$ \cite{Davies:2016jie} in $\gamma_{\mathrm{ns}}^{(3)}(N)$ have already been known for a long time.
In the general case of $n=3$, however, the possible basis elements typically outnumber by far the constraints in terms of the established low-$N$ moments of $\gamma_{ij}^{(n)}(N)$.

The salient point here is that the coefficients in the ansatz empirically come as relatively harmless fractions, made of modest integers modulo some powers of 2 and 3.
Therefore, we may try to find them using number theory, by viewing the established system of linear equations as a Diophantine system, which requires far less equations than unknowns, and applying to the matrix thus constructed the Lenstra-Lenstra-Lov\'{a}sz (LLL) algorithm \cite{Lenstra82factoringpolynomials} as implemented in the program package \texttt{fplll} \cite{fplll}.
This produces a matrix in which the rows are the solutions of the initial system of equations, but with the minimal Euclidean norms possible.
This method for the analytic reconstruction of the all-$N$ form of some anomalous dimension from its first few moments in $N$ was first proposed in the context of ${\mathcal N}=4$ SYM \cite{Velizhanin:2010cm} and was then successfully applied in ${\mathcal N}=4$ \cite{Velizhanin:2013vla,Marboe:2014sya,Marboe:2016igj,Kniehl:2020rip,Kniehl:2021ysp,Velizhanin:2021bdh,Kniehl:2023bbk,Kniehl:2024tvd} and $\mathcal{N}=2$ SYM \cite{Kniehl:2023bbk} and also in QCD \cite{Davies:2016jie,Moch:2017uml,Velizhanin:2012nm}.

Fortunately, the rapid-growth problem of function space dimensionality may be drastically mitigated in the flavor non-singlet sector by means of a generalization of the Gribov--Lipatov reciprocity relation, originally observed as $P_{ii}^{(0)}(x)=-xP_{ii}^{(0)}(1/x)$ or, equivalently, as the quasi-invariance of $\gamma_{ii}^{(0)}(N)$ under the mapping $N\to-1-N$ \cite{Gribov:1972rt}. 
In the case of $\gamma_{\mathrm{ns}}^{(n)}(N)$, the latter invariance property is only satisfied for the reciprocity respecting (RR) part $\mathcal{P}_{\mathrm{ns}}^{(n)}(N)$, which is determined by self-tuning \cite{Dokshitzer:2005bf,Dokshitzer:2006nm,Basso:2006nk},
\begin{equation}
\gamma(N)=\mathcal{P}(N-\gamma(N)-\beta(a_s)/a_s)\,,
\end{equation}
or, equivalently,
\begin{equation}
  \mathcal{P}(N)=\gamma(N+\mathcal{P}(N)+\beta(a_s)/a_s)\,,
  \label{eq:rr}
\end{equation} 
where
\begin{equation}
\mu^2\frac{\mathrm{d}a_s}{\mathrm{d}\mu^2}=
\beta(a_s)=-\sum_{n=0}^\infty b_n\,a_s^{n+2}\,,
\end{equation}
is the Gell-Mann--Low function of QCD, with $b_0=11/3C_A-2/3n_f$ \cite{Gross:1973id,Politzer:1973fx}, {\it etc.} 
Taylor expanding the rhs.\ of Eq.~\eqref{eq:rr} in $N$ and recursively substituting this equation into itself, one may formally express $\mathcal{P}$ entirely through $\gamma$ and its derivatives as 
\begin{equation}
\mathcal{P}=\gamma+\sum_{i=1}^\infty\frac{1}{(i+1)!}\,\frac{\mathrm{d}^i(\gamma+\beta/a_s)^{i+1}}{\mathrm{d}N^i}\,.
\label{eq:rr1}  
\end{equation}
Perturbative expansion of Eq.~\eqref{eq:rr1} yields
\begin{equation}
\mathcal{P}=a_s\gamma^{(0)}+\sum_{n=1}^\infty a_s^{n+1}(\gamma^{(n)}+\NP^{(n)})\,,
\label{eq:rr2}
\end{equation}
with
\begin{eqnarray}
\NP^{(1)}&=&(b_{0} + \gamma^{(0)})\dot\gamma^{(0)}\,,\nonumber\\
\NP^{(2)}&=&(b_{1} + \gamma^{(1)})\dot\gamma^{(0)}
+ (b_{0} + \gamma^{(0)}) (\dot\gamma^{(0)})^2 
\nonumber\\&&{}
+ \frac{1}{2}(b_{0}+\gamma^{(0)})^2 \ddot\gamma^{(0)}
+ (b_{0} + \gamma^{(0)}) \dot\gamma^{(1)}\,,
\nonumber\\
\NP^{(3)}&=&
(b_{1} + \gamma^{(1)}) (\dot\gamma^{(0)})^2
+ (b_{0} + \gamma^{(0)}) (\dot\gamma^{(0)})^3 
\nonumber\\&&{}
+(b_{0}b_{1} + b_{1} \gamma^{(0)} + b_{0} \gamma^{(1)} +  \gamma^{(0)} \gamma^{(1)}) \ddot\gamma^{(0)}
\nonumber\\
&&{}+ (b_{1} + \gamma^{(1)}) \dot\gamma^{(1)}
+ [b_{2} + \gamma^{(2)} + \frac{3}{2}(b_{0}+\gamma^{(0)})^2\ddot\gamma^{(0)}
\nonumber\\&&{}
+2(b_{0} + \gamma^{(0)}) \dot\gamma^{(1)}] \dot\gamma^{(0)} 
+ \frac{1}{6}(b_{0} + \gamma^{(0)})^3 \dddot\gamma^{(0)}
\nonumber\\&&{}
+ \frac{1}{2}(b_0+\gamma^{(0)})^2\ddot\gamma^{(1)}
+ (b_{0} +\gamma^{(0)}) \dot\gamma^{(2)}\,,
\label{eq:nrr}
\end{eqnarray}
where dots denote derivatives wrt.\ $N$, and so on.
In other words, we have uniquely decomposed $\gamma=\mathcal{P}-\NP$, and the non-RR part $\NP$ is, in each order, expressed in terms of $\gamma$ and its derivatives wrt.\ $N$ at lower orders.
The problem of analytically reconstructing $\gamma_{\mathrm{ns}}^{(n)}(N)$ for all $N$ is thus reduced to the same problem for $\mathcal{P}_{\mathrm{ns}}^{(n)}(N)$.
The great advantage of this resides in the observation that the function space thus collapses to the much smaller RR subspace of binomial harmonic sums (BHS's) \cite{Vermaseren:1998uu},
\begin{equation}
\BS_{m_1\ldots m_M}\!(N)\!=\!(-1)^N\!\sum_{i=1}^{N}\!(-1)^i\!\binom{N}{i}\!\!\binom{N+i}{i}\!S_{m_1,...,m_M}(i)\,,
\end{equation}
where $m_j\in\mathbb{N}$.
In fact, there are only $2^{w-1}$ such sums with weight $w$.
Furthermore, the denominators $1/N$ and $1/(N+1)$ only enter in the RR combination $\eta=1/[N(N+1)]$.
This suggests the ansatz
\begin{equation}
\mathcal{P}_{\mathrm{ns}}^{(n)}(N)=\sum_{w=0}^{2n+1}c_w\BS_{\vec{\mathbf{m}}_w}(N)
+\sum_{k=1}^{2n+1}\sum_{w=0}^{2n+1-k}c_{kw}\eta^k\BS_{\vec{\mathbf{m}}_w}(N)\,,
\end{equation}
with rational coefficients $c_w$ and $c_{kw}$, possibly multiplied by $\zeta_k$.

Notice that derivatives of $S_{\vec{\mathbf{m}}}(N)$ wrt.\ $N$ also generate terms involving $\zeta_k$ values.
Such terms unnecessarily pollute $\NP$ in Eq.~\eqref{eq:nrr} and feed into $\mathcal{P}$ via Eq.~\eqref{eq:rr2}, complicating the analytic reconstruction of the latter.
This may be neatly avoided by taking the derivatives in Eq.~\eqref{eq:nrr} to be incomplete, by dropping those spurious $\zeta_k$ terms, bearing in mind that they would anyway cancel in the combination $\gamma=\mathcal{P}-\NP$, being endowed with a simple structure, as in Eq.~\eqref{eq:zeta3} for $n=3$.

We now proceed with the analytic reconstruction of $\gamma_{\zeta_3}^{(3)}(N)$.
This is straightforward for the $n_f^3$ and $n_f^2$ contributions, and we find agreement with Refs.~\cite{Gracey:1994nn,Davies:2016jie}.
The residual color factors include $C_F^3n_f$, $C_F^2C_An_f$, $C_FC_A^2n_f$, $d_{44}^{\mathrm{RR}}n_f$, $C_F^4$, $C_F^3C_A$, $C_F^2C_A^2$, $C_FC_A^3$, $d_{44}^{\mathrm{RA}}$, where $d_{44}^{\mathrm{RR}}=d_F^{abcd}d_F^{abcd}/N_R=(n_c^2-1)(n_c^4-6n_c^2+18)/(96n_c^3)$ and $d_{44}^{\mathrm{RA}}=d_F^{abcd}d_A^{abcd}/N_R=(n_c^2-1)(n_c^2+6)/48$ stem from non-planar topologies.
We start with the fermionic contributions linear in $n_f$.
At first sight, the maximum weight is 3, namely, $2\times3+1$ minus 3 from $\zeta_3$ minus 1 from $n_f$, yielding the basis
\begin{eqnarray}
&&\{1,\eta,\BS_1,\eta^2,\eta\BS_1,\BS_2,\BS_{1,1},\eta^3,\eta^2\BS_1,\eta\BS_2,\eta\BS_{1,1},\BS_3,\BS_{2,1},
\nonumber\\
&&\qquad\BS_{1,2},\BS_{1,1,1}\}\,.
  \label{eq:basis}
\end{eqnarray}  
By analogy to the all-$N$ result in the large-$n_c$ limit \cite{Moch:2017uml}, however, we heuristically expect that Eq.~\eqref{eq:basis} has to be extended by $\{\eta^3\BS_1,\eta^2\BS_2,\eta^2\BS_{1,1}\}$, actually being of $w=4$, so that we have 18 basis functions altogether, whose coefficients can be successfully determined from the 8 available moments each for even and odd $N$ \cite{Velizhanin:2011es,Velizhanin:2014fua,Moch:2017uml} using the LLL algorithm \cite{Lenstra82factoringpolynomials,fplll}.
Since non-planarity sets on at $n=3$, reciprocity is respected for the
$d_{44}^{\mathrm{RR}}n_f$ term by itself.
However, the planar $C_F^3n_f$, $C_F^2C_An_f$, and $C_FC_A^2n_f$ terms all receive additional non-RR contributions, evaluated from Eq.~\eqref{eq:nrr} implemented with the incomplete derivative wrt.\ $N$.
%\footnote{%
%In general, derivatives of $S_{\vec{\mathbf{m}}}(N)$ wrt.\ $N$ also generate terms involving multiple zeta values.
%These may be safely dropped, in the sense of an incomplete derivative, keeping in mind the structure of Eq.~\eqref{eq:zeta3}.}

We now turn to the purely gluonic contributions, devoid of $n_f$.
Now, Eq.~\eqref{eq:basis} needs to be complemented by the $w=4$ basis functions
$\{\eta^4$, $\eta^3\BS_1$, $\eta^2\BS_2$, $\eta^2\BS_{1,1}$, $\eta\BS_3$, $\eta\BS_{2,1}$, $\eta\BS_{1,2}$, $\eta\BS_{1,1,1}$, $\BS_4$, $\BS_{3,1}$,
$\BS_{2,2}$, $\BS_{1,3}$, $\BS_{2,1,1}$, $\BS_{1,2,1}$, $\BS_{1,1,2}$, $\BS_{1,1,1,1}\}$, yielding 31 ones altogether.
In the $d_{44}^{\mathrm{RA}}$ term, the coefficients of the 8 BHS's with $w=4$ coincide with those of the non-planar universal anomalous dimension in $\mathcal{N}=4$ SYM \cite{Kniehl:2020rip,Kniehl:2021ysp,Kniehl:2024tvd} by the maximal-transcendentality principle \cite{Kotikov:2002ab,Kotikov:2004er}.
We are thus left with 23 basis functions, whose coefficients can be successfully determined using the LLL algorithm \cite{Lenstra82factoringpolynomials,fplll}.
The treatment of the planar contributions is too involved to be explained here, and we refer the interested reader to a forthcoming communication \cite{Kniehl2025:long}, where full details will be presented, including analytic expressions for $\gamma_{\zeta_3}^{(3)}(N)$ in SU($n_c$) color gauge theory and the respective parts $P_{\zeta_3}^{(3)\pm}(x)$ of $P_{\mathrm{ns}}^{(3)\pm}(x)$.
Notice that, in general, the inverse Mellin transformation of nested harmonic sums also generates $\zeta_k$ values.
In fact, $P_{\zeta_3}^{(3)\pm}(x)$, which multiply $\zeta_3$, turn out to contain $\zeta_2$, $\zeta_3$, and $\zeta_4$.

Our result for $\gamma_{\zeta_3}^{(3)}(N)$ passes all the checks mentioned above, against Refs.~\cite{Moch:2017uml,Gehrmann:2023iah,Beisert:2006ez,Bajnok:2008qj}.
Specifically, the large-$n_c$ limit is reached by putting $2C_F=C_A=n_c$, $d_{44}^{\mathrm{RA}}=n_c^3/96$, and $d_{44}^{\mathrm{RA}}=n_c^4/48$, and the result of $\mathcal{N}=4$ SYM is extracted by putting $C_F=C_A=n_c$, $d_{44}^{\mathrm{RA}}=n_c^4/24$, $n_f=0$ and retaining only the BHS's with $w=4$.

In real QCD, with $n_c=3$, the as-yet missing part of $\gamma_{\zeta_3}^{(3)}(N)$ reads 
\begin{widetext}
\begin{eqnarray}
\gamma_{\zeta_3}^{(3)}&=&
\frac{1024}{27}
\bigg[
-\frac{26307}{128}
+\frac{1}{144}(8111+20763\varepsilon)\eta
+\frac{477}{2}S_{1}
+\frac{5}{8}(10-53\varepsilon)D_1^2
+\frac{1}{48}(1721+5970\varepsilon)\eta^2
\nonumber\\&&{}
-\frac{1}{48}(15977-2586\varepsilon)\eta S_{1}
+\frac{25}{4}S_{2}
-\frac{1}{144}(31059-71\varepsilon)S_{-2}
-\frac{185}{4}\varepsilon D_1^3
+\frac{1}{288}(3691-2389\varepsilon)\eta^3
-10\varepsilon D_1^2S_{1}
\nonumber\\&&{}
-\frac{1}{48}(1963+810\varepsilon)\eta^2S_{1}
-\frac{1}{24}(7289+528\varepsilon)\eta S_{2}
-\frac{1}{144}(8928+2269\varepsilon)\eta S_{-2}
+\frac{1}{12}(7289+408\varepsilon)\eta S_{1,1}
\nonumber\\&&{}
+\frac{1}{8}(399+62\varepsilon)S_{-3}
-\frac{1}{8}(812+31\varepsilon)S_{3}
-\frac{1}{4}(351+62\varepsilon)S_{-2,1}
+\frac{81}{2}S_{1,-2}
-15\varepsilon D_1^4
+\frac{1}{24}(135-71\varepsilon)\eta^4
\nonumber\\&&{}
-20\varepsilon D_1^3S_{1}
+\frac{1}{48}(9217+1396\varepsilon)\eta^3S_{1}
+10D_1^2S_{-2}
-\frac{1}{24}(603+532\varepsilon)\eta^2S_{-2}
-\frac{1}{24}(6805-816\varepsilon)\eta^2S_{1,1}
\nonumber\\&&{}
+\frac{1}{48}(6805-1056\varepsilon)\eta^2S_{2}
-\frac{1}{8}(4+31\varepsilon)\eta S_{3}
-\frac{1}{2}(214+31\varepsilon)\eta S_{-2,1}
+\frac{1}{4}(120+31\varepsilon)\eta S_{-3}
+27\eta S_{1,-2}
-53S_{-3,1}
\nonumber\\&&{}
+44S_{-2,2}
-60S_{1,-3}
+S_{1,3}
+37S_{2,-2}
-85S_{3,1}
-68S_{-2,1,1}
+214S_{1,-2,1}
-54S_{1,1,-2}
-\frac{7}{6}S_{-4}
+\frac{119}{6}S_{4}
\nonumber\\&&{}
+\frac{133}{3}S_{-2,-2}
\bigg]
+\frac{512}{27}n_f
\bigg[
\frac{15241}{96}
+\frac{1}{48}(5819-360\varepsilon)\eta
-\frac{3463}{12}S_{1}
+\frac{5}{4}(25+2\varepsilon)D_1^2
+\frac{1}{24}(382-219\varepsilon)\eta^2
\nonumber\\&&{}
+\frac{1}{2}(46+7\varepsilon)\eta S_{1}
+\frac{35}{4}S_{-2}
+\frac{45}{4}S_{2}
+5(4+\varepsilon)D_1^3
+\frac{1}{12}(113-10\varepsilon)\eta^3
+40D_1^2S_{1}
-\frac{1}{2}(52-7\varepsilon)\eta^2S_{1}
+\frac{5}{2}\eta S_{-2}
\nonumber\\&&{}
+\frac{29}{2}\eta S_{2}
-69\eta S_{1,1}
-\frac{197}{6}S_{3}
+S_{-3}
-7S_{-2,1}
-5S_{1,-2}
+40S_{1,2}
+40S_{2,1}
-\frac{69}{2}\eta^2S_{2}
+69\eta^2S_{1,1}
-\frac{69}{2}\eta^3S_{1}
\bigg]\,,\qquad
\label{eq:new}
\end{eqnarray}
\end{widetext}
where $D_1=1/(N+1)$ and the argument $N$ has been dropped.
The missing part of $P_{\zeta_3}^{(3)+}(x)$, which is relevant for collisions with unpolarized hadrons, behaves for $x\to0$ as
\begin{eqnarray}
\lefteqn{P_{\zeta_3}^{(3)+}(x)=
\frac{4096}{243} \ln^3x+
%(\frac{9440}{81}-\frac{6592}{81}n_f)
\frac{32}{81}(295-206n_f)
\ln^2x+\frac{64}{81}}
\nonumber\\
&&\!\!\!\!{}\times[ 8320+ 10613 \zeta_2-(455 + 828 \zeta_2)n_f]
\ln x+ \frac{128}{243}(10523
\nonumber\\
&&\!\!\!\!{}+ 51918 \zeta_2 + 11532 \zeta_3) 
-\frac{32}{81} n_f (7279 + 3888 \zeta_2 + 1656 \zeta_3)\,,
\nonumber\\
\end{eqnarray}
and for $x\to1$ as
\begin{eqnarray}
\lefteqn{P_{\zeta_3}^{(3)+}(x)=
  - 1024 \zeta_2 \frac{\ln(1 - x)}{(1 - x)_+}
+ [\frac{27136}{3}
-768 \zeta_2  
}\nonumber\\&&\!\!\!\!{}
- 3072 \zeta_3 +
%n_f (-\frac{443264}{81} + \frac{21760}{27}\zeta_2) 
\frac{128}{81}\,n_f(-3463+510\zeta_2)
]\frac{1}{(1 - x)_+}
+ [\frac{23384}{3} 
\nonumber\\&&\!\!\!\!{}
-\frac{1049216}{243}\zeta_2
+ \frac{91264}{27}\zeta_3 
+ \frac{27328}{9}\zeta_4 
-
%n_f (\frac{243856}{81}
\frac{16}{81}\,n_f(15241
\nonumber\\&&\!\!\!\!{}
%+\frac{3520}{27}\zeta_2
%+ \frac{17536}{81}\zeta_3 ) 
+660\zeta_2+1096\zeta_3)
]\delta(1 - x)
+
%(1024 \zeta_2 - n_f \frac{20480}{27})
\frac{1024}{27}(27\zeta_2-20n_f)
\nonumber\\&&\!\!\!\!{}
\times
\ln(1-x)
-\frac{237824}{27} 
+ \frac{1792}{27}\zeta_2
+ 3072\zeta_3
\nonumber\\&&\!\!\!\!{}
+
%n_f(\frac{460544}{81}
%- \frac{21760}{27}\zeta_2)
\frac{256}{81}\,n_f(1799-255\zeta_2)
\,,
\label{eq:one}
\end{eqnarray}
where plus distributions $[d(x)]_+$ are defined as $\int_0^1\mathrm{d}x\,[d(x)]_+f(x)=\int_0^1\mathrm{d}x\,d(x)[f(x)-f(1)]$.
Notice that the extraordinary $\ln(1-x)/(1-x)_+$ term in Eq.~\eqref{eq:one} will cancel against a similar term in $P_{\mathrm{rat}}^{(3)+}(x)$.

\begin{figure}
\begin{center}
\includegraphics[width=0.47\textwidth]{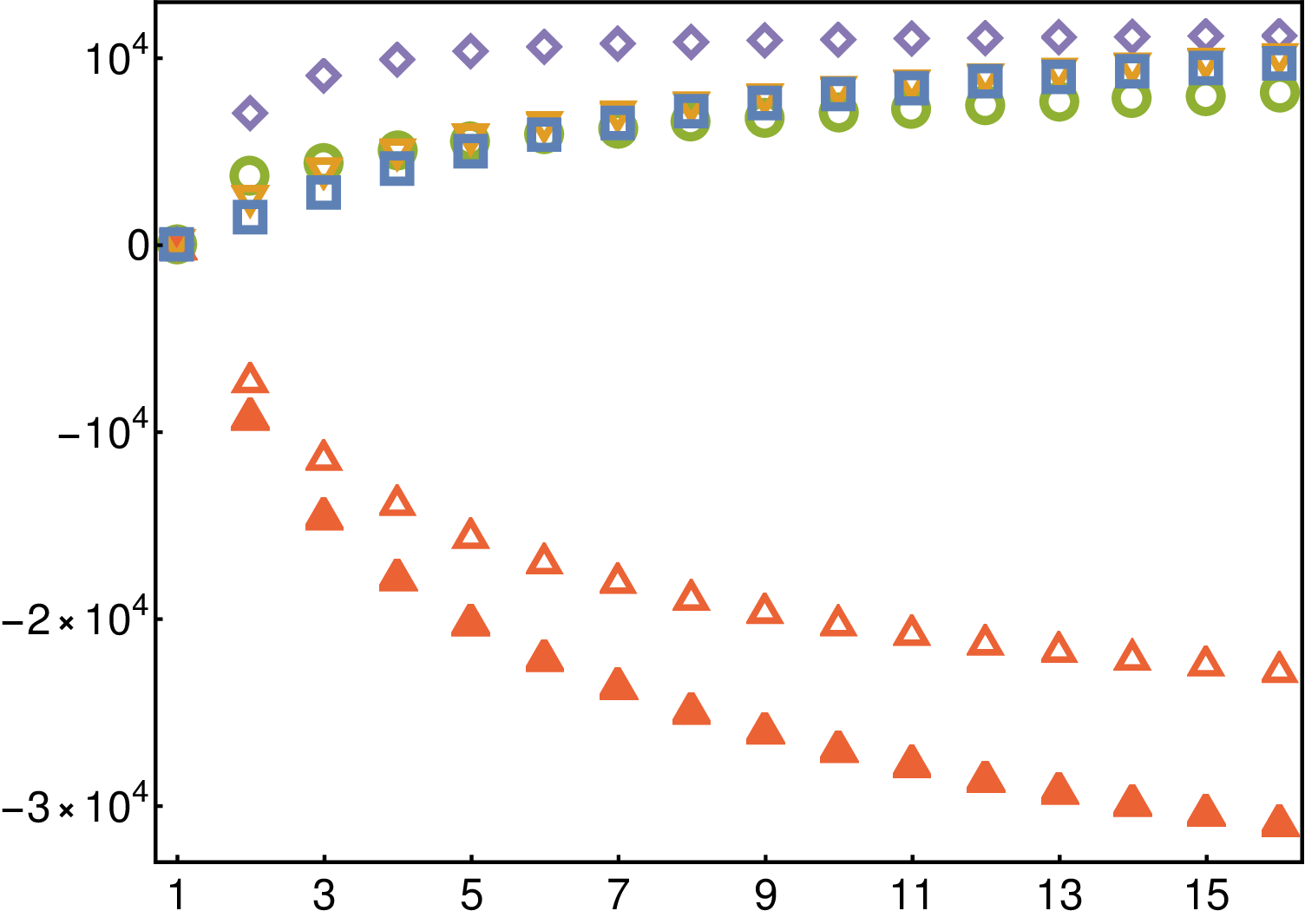}
\caption{$\gamma_{\mathrm{rat}}^{(3)}(N)$ (purple rhombs), $\zeta_3\gamma_{\zeta_3}^{(3)}(N)$ (red triangles), $\zeta_4\gamma_{\zeta_4}^{(3)}(N)$ (yellow triangles), $\zeta_5\gamma_{\zeta_5}^{(3)}(N)$ (blue squares) and $\gamma_{\mathrm{ns}}^{(3)}(N)$ (green circles) are shown for $n_c=3$ and $n_f=4$.
The new part of $\zeta_3\gamma_{\zeta_3}^{(3)}(N)$ (solid red triangles), without the $n_f^3$ \cite{Gracey:1994nn} and $n_f^2$ \cite{Davies:2016jie} contributions, is shown for comparison.}
\label{fig:moments}
\end{center}
\end{figure}

Our new results for $\gamma_{\zeta_3}^{(3)}(N)$ and $P_{\zeta_3}^{(3)\pm}(x)$ in QCD are numerically important.
This may be seen from Fig.~\ref{fig:moments}, where $\gamma_{\mathrm{ns}}^{(3)}(N)$ is broken down as in Eq.~\eqref{eq:zeta3} for $n_f=4$, appropriate for EIC energies.
In fact, the contribution from Eq.~\eqref{eq:new} is larger in magnitude than any other contribution shown.

We conclude with a brief lookout.
The completion of the transcendental part of $\gamma_{\mathrm{ns}}^{(3)}(N)$ in Eq.~\eqref{eq:zeta3}, achieved here, also benefits the analytic reconstruction of its rational part $\gamma_{\mathrm{rat}}^{(3)}(N)$ still left to be done.
This may be understood by observing that $\gamma_{\mathrm{rat}}^{(3)}(N)$ generates $\zeta_k$ values in the limits $N\to0$ and $N\to\infty$, which are to be taken to impose constraints from the generalized double-logarithmic equation \cite{Kirschner:1982qf,Kirschner:1983di,Velizhanin:2014dia} and the cusp anomalous dimension \cite{Henn:2019swt,vonManteuffel:2020vjv}, respectively.
Among the coefficients of the BHS's in the ansatz for the RR part $\mathcal{P}_{\mathrm{rat}}^{(3)}(N)$, those receiving a specific $\zeta_k$ factor in any such limit are now collectively subject to a respective constraint that is uniquely determined by the other inputs being all known.
As explained above, these BHS's have $w\le7$, and the non-RR part $\NP_{\mathrm{rat}}^{(3)}(N)$ is anyway fixed by known results from three loops and below.
Unfortunately, our present knowledge of $\gamma_{\mathrm{rat}}^{(3)}(N)$ for fixed values of $N$ is not yet rich enough for its analytic reconstruction to succeed.

{\it Note added:} After submission, S.-O. Moch informed us that he and his collaborators had independently obtained an all-$N$ result for $\gamma_{\zeta_3}^{(3)}(N)$ in SU($n_c$) that fully agrees with ours \cite{Moch:2025pri}.   

%%%%%%%%%%%%%%%%%%%%%%%%%%%%%%%%%%%%%%%%%%%%%%%%%%%%%%%%%%%%%%%%%%

%\subsection*{Acknowledgments}

This work was supported by the German Research Foundation DFG through Grant Nos.~KN~365/13-2 and 365/16-1.
%%%%%%%%%%%%%%%%%%%%%%%%%%%%%%%%%%%%%%%%%%%%%%%%%%%%%%%%%%%%%%%%%%

\end{document}